\documentclass[conference]{IEEEtran}

\usepackage{booktabs} 
\usepackage{glossaries}
\usepackage{makecell}
\usepackage{hyperref}
\usepackage{adjustbox}
\usepackage{multirow}
\usepackage{graphicx}
\usepackage{subcaption}
\usepackage{cite}

\usepackage{xcolor}  
\usepackage[colorinlistoftodos,prependcaption,textsize=footnotesize]{todonotes}
\usepackage{xargs}                      
\definecolor{OwnAzure}{HTML}{336699}
\definecolor{OwnCerulean}{HTML}{CAE2FE}
\definecolor{OwnOliveGreen}{HTML}{556B2F}
\newacronym{qos}{QoS}{Quality-of-Service}
\newacronym{sla}{SLA}{Service Level Agreement}
\newacronym{as}{AS}{autoscaler}
\newacronym{dag}{DAG}{Directed Acyclic Graph}
\newacronym{iot}{IoT}{Internet-of-Things}
\newacronym{bot}{BoT}{Bag-of-Tasks}
\newacronym{vm}{VM}{Virtual Machine}

\begin{document}
\title{Technical Report: A Trace-Based Performance Study of Autoscaling
\\Workloads of Workflows in Datacenters}

%
%

\author{\IEEEauthorblockN{Laurens Versluis}
	\IEEEauthorblockA{Vrije Universiteit Amsterdam\\
		l.f.d.versluis@vu.nl}
	\and
	\IEEEauthorblockN{Mihai Neac\c{s}u}
	\IEEEauthorblockA{Vrije Universiteit Amsterdam\\
		m.neacsu@atlarge-research.com}
	\and
	\IEEEauthorblockN{Alexandru Iosup}
	\IEEEauthorblockA{Vrije Universiteit Amsterdam\\
		a.iosup@vu.nl}
	}



%
%



\maketitle

\begin{abstract}
To improve customer experience, datacenter operators offer support for simplifying application and resource management. 
For example, running workloads of workflows on behalf of customers is desirable, but requires increasingly more sophisticated autoscaling policies, that is, policies that dynamically provision resources for the customer. 
Although selecting and tuning autoscaling policies is a challenging task for datacenter operators, so far relatively few studies investigate the performance of autoscaling for workloads of workflows. 
Complementing previous knowledge, in this work we propose the first comprehensive performance study in the field. 
Using trace-based simulation, we compare state-of-the-art autoscaling policies across multiple application domains, workload arrival patterns (e.g., burstiness), and system utilization levels. 
We further investigate the interplay between autoscaling and regular allocation policies, and the complexity cost of autoscaling.
Our quantitative study focuses not only on traditional performance metrics and on state-of-the-art elasticity metrics, but also on time- and memory-related autoscaling-complexity metrics.
Our main results give strong and quantitative evidence 
about previously unreported operational behavior, for example, that autoscaling policies perform differently across application domains and by how much they differ.

\end{abstract}

\section{Introduction}
\label{sct:introduction}

Many application domains of datacenter computing, from science to industrial processes to engineering, are based today on the execution of complex {\it workloads comprised of workflows}. 
To run such workloads in datacenters offering services as {\it clouds}, customers and providers must agree on a shared set of resource management and scheduling practices that automate the execution of many inter-dependent tasks, subject to \gls{qos} agreements.
In particular, they must agree on how to continuously acquire and release resources using {\it autoscaling} approaches, to lower operational costs for cloud customers and to increase resource utilization for cloud operators. 
Although many autoscaling approaches have been proposed, currently their behavior is not studied comprehensively. 
This is an important problem, which contributes to the low utilization of current datacenter (and especially cloud-based) environments, as low as 6\%--12\%~\cite{vasan2010worth, heinze2014auto, dougherty2012model}.
The state-of-the-art in analyzing the performance of datacenter autoscalers uses a systematic approach consisting of multiple metrics, statistically sound analysis, and various kinds of comparison tournaments~\cite{papadopoulos2016peas,ilyushkin2017experimental}, yet still lacks diversity in the application domain and insight into the interplay between the components of the autoscaling system.
Addressing these previously unstudied factors, in this work we propose a comprehensive study of autoscaling approaches. 


We focus this work on workloads of workflows running in datacenters.
Although the workloads of datacenters keep evolving, in our longitudinal study of datacenter workloads we have observed that many core properties persist over time~\cite{DBLP:journals/internet/IosupE11}. 
Workloads exhibit often non-exponential, even bursty~\cite{DBLP:journals/tpds/Li10}, arrival patterns, and non-exponential operational behavior (e.g., non-exponential runtime distributions).
Workflows are increasingly more common to use in datacenter environments, and are still rising in popularity across a variety of domains~\cite{wu2015workflow,ilyushkin2017experimental}.
Importantly, the characteristics of workflows vary widely across application domain, as indicated by the workflow size and per-task runtime reported for scientific~\cite{DBLP:journals/fgcs/JuveCDBMV13}, industrial processes~\cite{ma2017ananke}, and engineering domains~\cite{iosup2008grid}.

Managing workloads of workflows, subject to QoS requirements, is a complex activity~\cite{DBLP:journals/concurrency/RodriguezB17}. 
Due to the often complex structure of workflows, resource demands can change significantly over time, dynamically as the workflow tasks are executed and reflecting the inter-task dependencies.
Thus, designing
and tuning autoscaling techniques to improve resource utilization in the datacenter, while not affecting workflow performance, is non-trivial. 

Studying autoscalers is a challenging activity, where innovation could appear in
formulating new research questions about the laws of operation of autoscalers, or in creating adequate, reproducible experimental designs that reveal the laws experimentally. 
In this work, we propose the following new research questions: \emph{Does the application domain have an impact on the performance of autoscalers?}, \emph{What is the performance of autoscalers in response to significant workload bursts?}, \emph{What is the impact on autoscaling performance of the architecture of the scheduling system, and in particular of the (complementary) allocation policy?}, 
and \emph{What is the impact on autoscaling performance of the datacenter environment?}

We create a simulation-based experimental environment, using trace-based, realistic workloads as input. 
Our choice for this setup is motivated by pragmatic and methodological reasons. 
Currently, no analytical model based on either state-space exploration or non-state-space methods\footnote{Taxonomy introduced by Kishor Trivedi at WEPPE'17.} has been shown able to capture the complex workload-system interplay observed in practice for the context of this work. Such analytical models are hampered, e.g., by the lack of process stationarity (i.e., burstiness and dynamic workloads) and the ``curse of dimensionality'' stemming from sheer scale. 
Similarly, conducting real-world experiments, such as our own large-scale experiments~\cite{ilyushkin2017experimental} is also challenging in this context, due to the high cost and long duration of experiments, and the unreproducibility of experiments or statistical inconclusiveness of reported results~\cite{papadopoulos2016peas}.


In summary, our contribution in this work is four-fold:
\begin{enumerate}
	\item We are the first to investigate the impact on autoscaling performance of the application domain~(Section~\ref{sct:different_domain_experiment}). To this end, we experiment with workloads derived from real-world traces corresponding to scientific, engineering, and industrial domains.
	
	\item We analyze the behavior of the autoscalers when faced with sudden peaks in resource demand by using bursty, real-world workloads~(Section~\ref{sct:bursty_workload_experiment}). 
	
	\item We are the first to analyze the impact of the allocation policy on the performance of autoscalers~(Section~\ref{sct:different_allocation_policies_experiment}).
	
	\item We are the first to investigate the behavior of autoscalers running across a diverse set of datacenter environments~(Section~\ref{sct:utilization_experiment}). We specifically consider here the impact of datacenter utilization on autoscaling performance.
	
\end{enumerate}
\section{System Model}
\label{sct:model}\label{sec:model}

In this section, we describe a model of the datacenter environment considered in this work, focusing on autoscaling. 

\subsection{Model Overview}\label{sec:model:overview}

\begin{figure}[t]
	\includegraphics[width=\columnwidth]{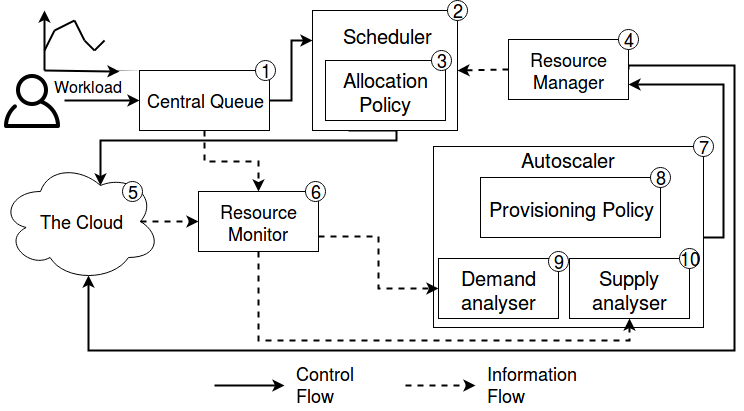}
	\caption{The cloud-based system model, focusing on resource autoscaling, including provisioning, and allocation.}
	\label{fig:architecture_dgsim+}
\end{figure}

We consider in this work the following model of a cloud-based datacenter environment that uses an autoscaling component to take decisions dynamically about resource acquisition and release.
As indicated by surveys of the field~\cite{vaquero2011dynamically, lorido2014review, ilyushkin2017experimental}, the state-of-the-art includes many autoscalers, designed for different environments, workloads, and other design constraints. 
In contrast, our model matches the operations we have observed in practice in 
(i) {\it public cloud environments}, such as Amazon AWS and Microsoft Azure, 
(ii) {\it semi-public clouds} servicing industries with strict security and process restrictions, such as the datacenters operated by the Dutch company Solvinity for the financial industry and government public-services, 
and 
(iii) {\it private}, albeit {\it multi-institutional}, {\it clouds} such as the research-cloud DAS~\cite{bal2016medium}.





Figure~\ref{fig:architecture_dgsim+} depicts an overview of our model.
The workloads is comprised of workflows, which are submitted by users to the \textit{central queue} of the datacenter~(component 1 in Figure~\ref{fig:architecture_dgsim+}).
The transient set of machines available to users forms the \textit{cloud}~(5).
An user-aware \textit{scheduler} calls an \textit{allocation policy}~(3) to place the queued workflows onto machines available in the cloud.
In parallel with the operation of the scheduler, the \textit{resource monitor}~(6) monitors periodically the utilization of each {\it active}, that is, allocated {\it machine}, and the state of the central queue. 
Starting from the monitoring information, and possibly also using historical data, the \textit{autoscaler} component~(7) periodically analyzes the demand (9) and the supply (10) of the dynamic system. 
Through the \textit{resource manager}~(4), the autoscaler can issue (de)allocation commands, to dynamically obtain or release resources according to the defined \textit{provisioning policy}~(8). Matching the state-of-the-art in such policies~\cite{papadopoulos2016peas,ilyushkin2017experimental}, the provisioning policies we consider in this work are periodic and not event-based.

The periods used by the resource monitor and by the autoscaler evaluation, that is, the\textit{monitoring interval} and the \textit{autoscaling interval}, respectively, are adjusted through configuration files.

\subsection{Workflows}

A workflow is a set of tasks with precedence constraints between them.
When all precedence constraints of a task have been fulfilled, it can be executed. 
The concept of a workflow fits both compute-intensive and data-intensive (i.e., for dataflows) tasks, and the precedence constraints frequently express data dependencies between tasks.

In this work, we use the common formalism of \glspl{dag} to model workflows.
In this formalism, tasks are modeled as vertices, and precedence constraints are modeled as edges.
\glspl{dag} are used across a variety of application domains, in science~\cite{DBLP:journals/fgcs/JuveCDBMV13}, industrial processes~\cite{ma2017ananke}, and engineering~\cite{iosup2008grid}. 
\glspl{dag} also fit well web hosting~\cite{wu2015workflow} and multi-tier web applications~\cite{urgaonkar2005dynamic}. For big data workloads, MapReduce workflows are frequently modeled as simple \glspl{dag}, where mappers and reducers are modeled as nodes, and the data dependencies between mappers and reducers are modeled as links in the graph; Spark-like dataflows are also common.


\begin{table*}[!htb]
	\centering
	\caption{The design and setup of our experiments. In bold, the distinguishing features of each experiment. (AS $=$ autoscaling.)}
	\label{tbl:experimental_setup}
	\begin{tabular}{|r||c|r|r|c|c|c|c|c|c|}
		\hline
		\makecell{Experiment \\Focus (Section)} & Workload      & \# Clusters  		& \makecell{\# Resources \\per cluster} & \makecell{AS \\policies}       & \makecell{Allocation \\Policies} & \makecell{Workflow \\Metrics} & \makecell{Autoscaler \\Metrics} & \makecell{Scale \\Metrics}               \\ \hline \hline
		Domain ($\S$\ref{sct:different_domain_experiment})      & \textbf{T1, T2, T3}    & 50                   & 70             & All               & FillworstFit    & -          & \textbf{All} & -                          \\ \hline
		Bursty ($\S$\ref{sct:bursty_workload_experiment})       	& \textbf{T2 (dupl), T4} & \textbf{Variable}                   & 70             & All               & FillworstFit    & \textbf{NSL}        & -          & -                           \\ \hline
		Allocation ($\S$\ref{sct:different_allocation_policies_experiment})       & T4            & 50                   & 70             & All               & \textbf{All}             & -          & \textbf{Supply}     & -                           \\ \hline
		Utilization ($\S$\ref{sct:utilization_experiment})      & T3            & \textbf{Variable}             & 70             & All               & FillworstFit    & NSL        & All & -                           \\ \hline
	\end{tabular}
\end{table*}

\subsection{Provisioning Policies}
\label{ssct:as_policies}

We consider in this work two classes of autoscalers: (i) general autoscalers that are agnostic to workflows, and (ii) autoscalers that exploit knowledge and structure of workloads of workflows. All policies used in this work have already been used in a prior studies in the community~\cite{papadopoulos2016peas,ilyushkin2017experimental} and provide a representative set of general and workflow-aware policies. They also cover a variety of strategies, based on different optimization goals such as throughput, (estimated) level of parallelism, and queue size.

We now describe the provisioning policies we study in this work, in turn:

\vspace*{-0.15cm}
\subsubsection{Reg} 
employs a predictive component based on second-order regression to compute future load~\cite{iqbal2011adaptive}. 
This policy works as follows: when the system is {\it underprovisioned}, that is, the capacity is lower than the load, \textit{Reg} takes scale-up decisions through a reactive component that behaves similarly to the \textit{React} policy (described in the following). 
When the system is {\it overprovisioned}, \textit{Reg} takes scale-down decisions based on its predictive component, which in turn uses workload history to predict predict future demand. 

\vspace*{-0.15cm}
\subsubsection{Adapt} 
autonomously changes the number of virtual machines allocated to a service, based on both monitored load-changes and predictions of future load~\cite{ali2014measuring}.
The predictions are based on observed changes in the request rate, that is, the slope of the workload demand curve. 
\textit{Adapt} features a controller aiming to respond to sudden changes in demand, and to not release resources prematurely.
The latter strategy aims to reduce oscillations in resource availability.

\vspace*{-0.15cm}
\subsubsection{Hist} 
focuses on the dynamic demands of multi-tier internet applications~\cite{urgaonkar2005dynamic}.
This policy uses a queueing model to predict future arrival rates of requests based on hourly histograms of the past arrival-rates.
\textit{Hist} also features a reactive component that aims to handle sudden bursts in network requests, and to self-correct errors in the long-term predictions.

\vspace*{-0.15cm}
\subsubsection{React Policy} 
takes resource provisioning decisions based on the number of active connections~\cite{Chieu2009DynamicSO}. 
\textit{React} first determines the number of resource instances below or above a given threshold. New instances are provisioned if all resource instances have an utilization rate above the threshold. 
If there are underutilized instances and at least one instance has no active connections (so, the instance is not in use), the idle instance is shutdown. 

\vspace*{-0.15cm}
\subsubsection{ConPaaS} 
is designed to scale web applications at fixed intervals, based on observed throughput~\cite{fernandez2014autoscaling}.
Using several time series analysis techniques such as Linear Regression, Auto Regressive Moving Average (ARMA), and Multiple Regression, \textit{ConPaaS} predicts future demand, and provisions resources accordingly.

\vspace*{-0.15cm}
\subsubsection{Token} 
is designed specifically to autoscale for workflows~\cite{DBLP:conf/ccgrid/IlyushkinGE15}.
By using structural information from the DAG structure, and by propagating execution tokens in the workflow, this policy estimated the level of parallelism and derives the (lack of) need for resources.
%
%

\vspace*{-0.15cm}
\subsubsection{Plan} 
 predicts resource demand based on both workflow structure (as \textit{Token} also does) and on task runtime estimation~\cite{ilyushkin2017experimental}. 
\textit{Plan} constructs a partial execution plan for tasks that are running or waiting in the queue, considering task placement only for the next autoscaling interval. 
The resource estimation is derived by also considering the number of resources that already have tasks assigned to them; \textit{Plan} places tasks on unassigned resources using a first-come, first-served order (FCFS) allocation policy.

%

\begin{table}[!htb]
	\centering
	\caption{The four types of workloads and their characteristics. (W = number of workflows.)}
	\label{tbl:workloads}
	\begin{tabular}{|c||l|l|r|r|}
		\hline
		ID & Source         & Domain      & W & \# Tasks \\ \hline\hline
		T1   & SPEC Cloud Group \cite{ilyushkin2017experimental} & Scientific  & 200 & 13,876   \\ \hline
		T2   & Chronos \cite{ma2017ananke} & Industrial  & 1,024 & 3,072     \\ \hline
		T3   & Askalon EE \cite{iosup2008grid}   & Engineering & 757          & 45,786   \\ \hline
		T4   & Askalon EE2 \cite{iosup2008grid} & Engineering & 3,551        & 122,105  \\ \hline
	\end{tabular}
\end{table}

\section{Experimental Design}
\label{sct:experiments}\label{sed:exp}


In this section we present our experimental design.
We elaborate on our workloads, metrics and other variables used to evaluate the performance of the autoscalers introduced in Section~\ref{ssct:as_policies}.
We introduce our experimental setup in Section~\ref{ssct:experimental_setup} and summarize it in Table~\ref{tbl:experimental_setup}.
All these components are implemented in our simulator, which is used to conduct the outlined experiments.

\subsection{Experiments and Workloads}
\label{ssct:experimental_setup}\label{sed:exp:setup}

\begin{figure}[!htb]
	\includegraphics[height=6cm, width=\columnwidth]{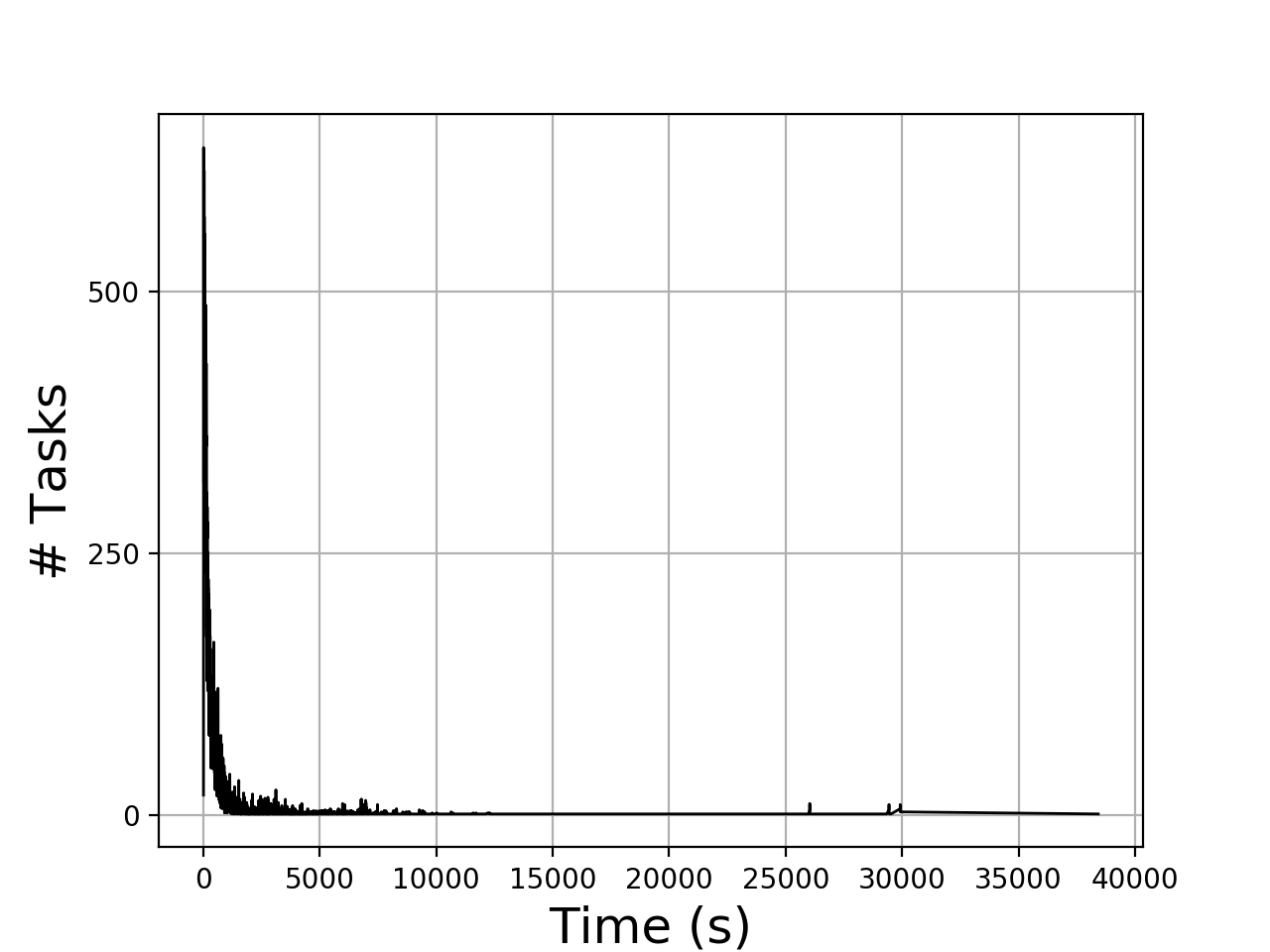}
	\caption{The arrival rate of Askalon EE2 (T4).}
	\label{fig:arrival_pattern_askalon_ee2}
\end{figure}

In total, this work uses four different workloads, listed in Table~\ref{tbl:workloads}.
Each workload consists of multiple workflows.
Each workflow consists of multiple tasks, with dependency constraints between them.
The required amount of CPUs and runtime per task are known a priori.
Besides T1, all other workloads are real-world traces taken from either production environments or scientific clusters.
T2 is derived from the Chronos production environment at Shell.
This workload contains several workflows assigned to three distinct levels.
A level is dependent on the previous, forming a chain.
We modeled this structure by chaining the workflows in our simulations.
T3 and T4 are traces from the Askalon cluster.
Both T3 and T4 feature a burst at the start of the workload.
Figure~\ref{fig:arrival_pattern_askalon_ee2} shows the arrival pattern of tasks
for T4.

We perform four experiments in total.
In the first experiment we run the Chronos workload (T1) and Askalon EE (T3) using an industrial setup used by the state-of-the-art and a setup derived from this industrial setup.
The second experiment focuses on bursts in workload.
Some of the autoscalers feature a reactive component, capable of correcting an autoscaler in case errors accumulate or to change to sudden changes in task arrival rates.
With this experiment we aim to observe how well the autoscalers react to such an increase in resource demand.
In the third experiment we measure the impact of different allocation policies on the behavior and performance of the autoscalers.
In the fourth experiment we investigate differences between autoscalers when running in different environments by changing changing the average resource utilization given a workload.

In all experimental setups, the simulator starts with all sites allocated, the autoscaler scales after its first evaluation period, also referred to as \textit{service rate} \cite{ilyushkin2017experimental}.
The service rate is set to 30 seconds for all experiments.
The resource manager receives real-time updates of resource usage of each site.
We do not include file transfer times or site boot-up times in any of the experiments.
By default, the simulator schedules tasks using the FillworstFit policy.
FillworstFit selects the site with the most available resources and assigns as much tasks as possible to this site before selecting another in a first come first serve order.
Matching the state-of-the-art comparison \cite{ilyushkin2017experimental}, clusters are only deallocated when they are idle, i.e. no tasks are being executed.
At least one cluster is kept running at all times.

\subsection{Metrics}\label{sed:exp:metrics}

Measuring the performance of a system and its components can be done at different levels.
In this work we use both well established and state-of-the-art metrics divided in three categories: workflow-level, autoscaler-level, and scale-level metrics.
Table~\ref{tbl:metrics} provides an overview of all metrics used in this work.

\begin{table}
	\caption{The metrics used in this work, grouped by level.}
	\label{tbl:metrics}
\begin{tabular}{|c|l|}
	\multicolumn{2}{l}{\textbf{Workflow-level metrics}} \\ \hline
	M & Makespan \\ \hline
	W & Wait time \\ \hline
	R & Reponse time \\ \hline
	NSL & Normalized schedule length \\ \hline
	CP & Critical path length \\ \hline
	S & Slowdown \\ \hline
	\multicolumn{2}{l}{\textbf{Autoscaler-level metrics}} \\ \hline
	$A_U$ & Accuracy underprovisioning \\ \hline
	$A_O$ & Accuracy overprovisioning \\ \hline
	$\bar{A}_U$ & Normalized accuracy underprovisioning \\ \hline
	$\bar{A}_O$ & Normalized accuracy overprovisioning \\ \hline
	$T_U$ & Time underprovisioned \\ \hline
	$T_O$ & Time overprovisioned \\ \hline
	k & Average fraction of time of overprovisioning trends \\ \hline
	k' & Average fraction of time of underprovisioning trends \\ \hline
	$M_U$ & Average number of idle resources \\ \hline
	$\bar{V}$ & Average number of resources \\ \hline
	$\bar{h}$ & Average accounted CPU hours per \gls{vm} \\ \hline
	$\bar{C}$ & Average charged CPU hours per \gls{vm} \\ \hline
	\multicolumn{2}{l}{\textbf{Scale-level metrics}} \\ \hline
	I & Amount of instructions \\ \hline
	D & Amount of data items in memory \\ \hline
	
\end{tabular}
\end{table}

\subsubsection{Workflow-level Metrics}

Workflow-level metrics measure the performance of a system at the workflow level.
These metrics define the impact an autoscaler has on workflow execution speed.

In this work we consider the following five workflow-level metrics:


The makespan (M) of a workflow is defined as the time elapsed from the start of its first task until the completion of its last task. 
This represents the total execution time of the workflow and it includes the execution time of its tasks, the data transfer time and the time spent in the scheduling queues from the moment the first entry task starts its execution up until the last exit task is completed.
The wait time (W) is the time elapsed from when a task becomes eligible to execute until the moment it begins executing on resources. We only consider the wait time of the workflow which is the time elapsed between the arrival and the start of execution of a workflow's first task.
The response time (R) of a workflow is the total time a workflow stays in the system. It is the summation of the makespan and wait time of a workflow.
The Normalized Schedule Length (NSL) of a workflow is its response time normalized by the critical path length. 
The critical path length (CP) is the longest path that affects workflow completion time and is computed using known runtime and data transfer time estimates. Therefore, the Normalized Schedule Length allows us to compare the workload response time to the ideal case where most time is spent on the critical path and no delays are incurred.
The slowdown (S) of a workflow is its response time in a system that runs an autoscaler, normalized by its response time in a system of the same size that does not run an autoscaler and keeps the number of resources constant during its execution. 
This allows us to see how much the autoscaler slows down the system due to resource throttling.

\subsubsection{Autoscaler-level Metrics}
\label{sssct:autoscaler-level_metrics}

Autoscaler-level metrics define how well an autoscaler is performing.
Ilyushkin et al.~\cite{ilyushkin2017experimental} define a large set of metrics which we adopt in this work.


The average demand is the amount of resources that are required to uphold a Service Level Objective (SLO) at each autoscale step divided by the execution time.
Similarly, the average supply is the average amount of resources provisioned over the execution time. In our model, we count as supplied any resource that is in a running or shutting down state.
The accuracy underprovisioning metric ($A_U$) describes the average amount of resources that are under-provisioned during the execution of our simulation. 
This is computed by summing the number of resources that the autoscaler provides too low in relation to the momentary demand divided by the total number of resources that can be available during the experiment.
Similarly, the accuracy overprovisioning metric ($A_O$) describes the average amount of resources that exceed the demand during the execution of our simulation.

The accuracy metrics however, can prove to be unfit if the number of resources under- or over-provisioned varies considerably over time. 
By normalizing the accuracy under- and overprovisioning metrics by the momentary resource demand, respectively supply, describes more accurately how much the system is under- or over-provisioned. 
Hence, we use normalized accuracy overprovisioning~($\bar{A}_O$) and normalised accuracy underprovisioning~($\bar{A}_U$).
Different from Ilyushkin et al., we define $\bar{A}_O$ as $\frac{1}{T} \sum_{t=1}^{T} \frac{(s_t - d_t)^{+}}{max(s_t, \epsilon)}$ to obtain a value between 0 and 1, where $T$ is the total elapsed time of the experiment, $s_t$ is the supply at time $t$, $d_t$ the demand at time $t$, $(a)^+ := max(a, 0)$, and $\epsilon = 1$ as per state-of-the-art~\cite{ilyushkin2017experimental}.

The number of resources provisioned needs to follow the demand curve as accurately as possible. 
Overprovisioning will result in idle resources that do not speed up workload processing and incur additional cost. 
The ideal autoscaler provisions resources so that the system is executing all eligible tasks without having any idle resources. 
The average number of idle resources ($M_U$) during the execution captures this property.
Time underprovisioned ($T_U$) and over-provisioned ($T_O$) measure the fraction of time a system is over- or underprovisioned.
These metrics can convey if the system is constantly not following the resource demand curve or just a few times with large deviations.
The average amount of resources allocated ($\bar{V}$) provides insight into the costs of an autoscaler.
The average accounted CPU hours per \gls{vm} ($\bar{h}$) measures the amount of hours a \gls{vm} was used i.e., the effective CPU hours used.
The average charged CPU hours ($\bar{C}$) measures the total amount of CPU hours a \gls{vm} is charged.

\subsubsection{Scale-level metrics}

Scale-level metrics define the ability of an algorithm or policy to scale with the workload and environment.
In real-world settings autoscaling must take place in the order of seconds.
It is therefore important that under heavy load or at a large scale an autoscaler still performs well.
In our study, we perform a worst-case analysis by looking at the big-O notation for each autoscaler.
By counting the number of instructions (I) an autoscaler performs throughout the execution of a workload, it can be measured if certain autoscalers require significantly more computation.
Similarly, by keeping track of the amount of data items (D) in memory during the execution of a workload, we observe how well autoscalers scale with respect to workload and environment.



\subsection{Implementation Details}
\label{sed:exp:implementation}

We implement the model from Section~\ref{sec:model} as a simulation prototype in the OpenDC collaborative datacenter simulation project~\cite{conf/ispdc/IosupABBENOTVV17}; the resulting prototype, OpenDC.workflow, uses and extends significantly earlier code from the DGSim project~\cite{iosup2008dgsim}.

The prototype is implemented in Python 2.7 and features all components highlighted in Section~\ref{sec:model:overview}.
The implementation is modular, allowing components such as autoscaling and allocation policies to be swapped by configuration.
The prototype is open-source\footnote{The project will be available online before the conference.} and available at \url{www.github.com/atlarge-research/opendc/autoscaling-prototype}.

\section{Application Domain Experiment}
\label{sct:different_domain_experiment}

\begin{table*}[] 
	\centering
	\caption{The elasticity results, per workload, per autoscaler (AS), using a {\it static} infrastructure.}
	\label{tbl:workload_domain_experimenSPEC}
	\begin{tabular}{|c|c||r|r|r|r|r|r|r|r|r|r|r|r|}
		\hline
		AS & Workload & $A_U$ & $A_O$ & $\bar{A}_U$ & $\bar{A}_O$ & $T_U$ & $T_O$ & k & k' & $M_U$ & $\bar{V}$ & $\bar{h}$ & $\bar{C}$ \\ \hline
		\multirow{3}{*}{Reg} & Chronos & 3.3 & 6.9 & 21.9 & 33.5 & 46.2 & 50.8 & 0.0 & 0.0 & 4.2 & 420.0 & 432.0 & 2.6\\
		& Askalon EE & 63.3 & 15.0 & 18.4 & 47.1 & 25.8 & 56.6 & 2.0 & 2.0 & 15.6 & 1,491.8 & 1,534.4 & 42.9\\
		& SPEC & 0.1 & 3.4 & 2.3 & 65.1 & 5.0 & 93.7 & 2.3 & 0.6 & 4.4 & 176.9 & 182.0 & 40.3\\ \hline
		\multirow{3}{*}{Hist} & Chronos & 3.3 & 6.9 & 21.9 & 33.5 & 46.2 & 50.8 & 0.0 & 0.0 & 4.2 & 420.0 & 432.0 & 2.6\\
		& Askalon EE & 63.7 & 71.2 & 16.5 & 71.2 & 24.0 & 75.0 & 19.0 & 0.0 & 72.7 & 3,466.2 & 3,565.2 & 99.0\\
		& SPEC & 0.1 & 5.0 & 1.5 & 70.2 & 5.5 & 94.4 & 3.6 & 0.0 & 6.0 & 235.0 & 241.8 & 53.6\\ \hline
		\multirow{3}{*}{ConPaaS} & Chronos & 3.3 & 6.9 & 21.9 & 33.5 & 46.2 & 50.8 & 0.0 & 0.0 & 4.2 & 420.0 & 432.0 & 2.6\\
		& Askalon EE & 63.7 & 51.4 & 16.5 & 70.7 & 24.0 & 75.0 & 19.0 & 0.0 & 52.0 & 2,773.6 & 2,852.9 & 79.2\\
		& SPEC & 0.8 & 1.3 & 10.6 & 46.4 & 24.8 & 74.5 & 3.6 & 0.0 & 1.9 & 91.3 & 93.9 & 20.8\\ \hline
		\multirow{3}{*}{React} & Chronos & 3.3 & 6.9 & 21.9 & 33.5 & 46.2 & 50.8 & 0.0 & 0.0 & 4.2 & 420.0 & 432.0 & 2.6\\
		& Askalon EE & 63.7 & 15.3 & 17.5 & 55.4 & 25.0 & 65.0 & 10.0 & 1.0 & 15.8 & 1,508.3 & 1,551.4 & 43.1\\
		& SPEC & 0.1 & 3.4 & 1.8 & 65.3 & 3.8 & 94.7 & 2.0 & 0.8 & 4.5 & 180.5 & 185.7 & 41.1\\ \hline
		\multirow{3}{*}{Token} & Chronos & 0.0 & 48.0 & 0.0 & 78.0 & 0.0 & 96.9 & 0.0 & 0.0 & 49.3 & 1,977.2 & 2,033.7 & 12.2\\
		& Askalon EE & 63.7 & 16.4 & 16.5 & 65.8 & 24.0 & 75.0 & 19.0 & 0.0 & 17.0 & 1,548.2 & 1,592.5 & 44.2\\
		& SPEC & 0.0 & 69.1 & 0.0 & 95.1 & 0.0 & 99.9 & 3.6 & 0.0 & 70.2 & 2,481.2 & 2,552.1 & 565.4\\ \hline
		\multirow{3}{*}{Adapt} & Chronos & 3.3 & 6.9 & 21.9 & 33.5 & 46.2 & 50.8 & 0.0 & 0.0 & 4.2 & 420.0 & 432.0 & 2.6\\
		& Askalon EE & 63.7 & 15.6 & 17.5 & 55.3 & 25.0 & 65.0 & 10.0 & 1.0 & 16.2 & 1,520.2 & 1,563.6 & 43.4\\
		& SPEC & 0.3 & 2.2 & 5.2 & 52.8 & 15.9 & 82.1 & 2.0 & 0.8 & 3.1 & 131.7 & 135.4 & 30.0\\ \hline
		\multirow{3}{*}{Plan} & Chronos & 3.3 & 6.9 & 21.9 & 33.5 & 46.2 & 50.8 & 0.0 & 0.0 & 4.2 & 420.0 & 432.0 & 2.6\\
		& Askalon EE & 63.3 & 15.5 & 18.4 & 47.3 & 25.8 & 56.6 & 2.0 & 2.0 & 16.0 & 1,508.4 & 1,551.5 & 43.4\\
		& SPEC & 0.4 & 1.8 & 6.4 & 50.9 & 17.6 & 79.1 & 0.9 & 0.9 & 2.6 & 115.4 & 118.7 & 26.3\\ \hline
		
	\end{tabular}
\end{table*}

\begin{table*}[] 
	\centering
	\caption{The elasticity results, per workload, per autoscaler (AS), using a {\it workload-scaled} infrastructure.}
	\label{tbl:workload_domain_experiment2}
	\begin{tabular}{|c|c||r|r|r|r|r|r|r|r|r|r|r|r|}
		\hline
		AS & Workload & $A_U$ & $A_O$ & $\bar{A}_U$ & $\bar{A}_O$ & $T_U$ & $T_O$ & k & k' & $M_U$ & $\bar{V}$ & $\bar{h}$ & $\bar{C}$ \\ \hline
		\multirow{3}{*}{Reg} & Chronos & 70.7 & 17.6 & 27.8 & 17.6 & 45.9 & 50.5 & 0.0 & 0.0 & 36.0 & 202.3 & 3,467.9 & 21.0\\
		& Askalon EE & 210.0 & 16.1 & 33.1 & 45.6 & 40.7 & 53.6 & 2.0 & 2.0 & 16.7 & 1,312.4 & 2,109.2 & 59.0\\
		& SPEC & 53.9 & 39.7 & 15.4 & 39.7 & 33.4 & 65.3 & 2.4 & 0.5 & 71.6 & 69.4 & 3,570.5 & 791.0\\ \hline
		\multirow{3}{*}{Hist} & Chronos & 70.7 & 17.6 & 27.8 & 17.6 & 45.9 & 50.5 & 0.0 & 0.0 & 36.0 & 202.3 & 3,467.9 & 21.0\\
		& Askalon EE & 211.5 & 56.2 & 31.4 & 56.2 & 39.0 & 60.0 & 7.0 & 0.0 & 57.8 & 2,218.3 & 3,565.2 & 99.0\\
		& SPEC & 53.9 & 40.2 & 14.9 & 40.2 & 33.0 & 66.0 & 2.6 & 0.0 & 72.5 & 70.0 & 3,597.6 & 797.0\\ \hline
		\multirow{3}{*}{ConPaaS} & Chronos & 70.7 & 17.6 & 27.8 & 17.6 & 45.9 & 50.5 & 0.0 & 0.0 & 36.0 & 202.3 & 3,467.9 & 21.0\\
		& Askalon EE & 211.5 & 50.4 & 31.4 & 55.8 & 39.0 & 60.0 & 7.0 & 0.0 & 51.0 & 2,088.1 & 3,355.9 & 93.2\\
		& SPEC & 52.9 & 40.2 & 14.9 & 40.2 & 32.9 & 66.0 & 2.6 & 0.0 & 72.5 & 70.0 & 3,597.6 & 797.0\\ \hline
		\multirow{3}{*}{React} & Chronos & 70.7 & 17.6 & 27.8 & 17.6 & 45.9 & 50.5 & 0.0 & 0.0 & 36.0 & 202.3 & 3,467.9 & 21.0\\
		& Askalon EE & 211.5 & 16.1 & 32.4 & 47.6 & 40.0 & 56.0 & 4.0 & 1.0 & 16.7 & 1,319.9 & 2,121.3 & 58.9\\
		& SPEC & 54.8 & 38.9 & 15.7 & 38.9 & 33.7 & 64.4 & 1.6 & 0.8 & 70.8 & 68.8 & 3,538.9 & 784.0\\ \hline
		\multirow{3}{*}{Token} & Chronos & 70.7 & 17.6 & 27.8 & 17.6 & 45.9 & 50.5 & 0.0 & 0.0 & 36.0 & 202.3 & 3,467.9 & 21.0\\
		& Askalon EE & 211.5 & 16.8 & 31.4 & 51.4 & 39.0 & 60.0 & 7.0 & 0.0 & 17.4 & 1,335.3 & 2,146.1 & 59.6\\
		& SPEC & 53.4 & 40.3 & 14.8 & 40.3 & 32.7 & 66.1 & 2.6 & 0.0 & 72.5 & 70.0 & 3,597.6 & 797.0\\ \hline
		\multirow{3}{*}{Adapt} & Chronos & 70.7 & 17.6 & 27.8 & 17.6 & 45.9 & 50.5 & 0.0 & 0.0 & 36.0 & 202.3 & 3,467.9 & 21.0\\
		& Askalon EE & 211.5 & 17.6 & 32.4 & 47.5 & 40.0 & 56.0 & 4.0 & 1.0 & 18.2 & 1,353.6 & 2,175.4 & 60.4\\
		& SPEC & 54.9 & 38.8 & 15.7 & 38.8 & 33.6 & 64.6 & 1.6 & 0.8 & 70.9 & 68.8 & 3,538.9 & 784.0\\ \hline
		\multirow{3}{*}{Plan} & Chronos & 70.7 & 17.7 & 27.8 & 17.7 & 45.9 & 50.5 & 0.0 & 0.0 & 36.0 & 202.3 & 3,467.9 & 21.0\\
		& Askalon EE & 210.0 & 17.8 & 33.1 & 45.9 & 40.7 & 53.6 & 2.0 & 2.0 & 18.4 & 1,350.0 & 2,169.6 & 60.7\\
		& SPEC & 54.2 & 38.1 & 15.7 & 38.1 & 33.7 & 63.6 & 0.9 & 0.9 & 70.0 & 68.2 & 3,507.3 & 777.0\\ \hline
		
	\end{tabular}
\end{table*}

The aim of this experiment is to answer our research question \emph{Does the application domain have an impact on the performance of autoscalers?}, by running workloads from distinct domains.
In the state-of-the-art comparison study conducted by Ilyushkin et al. \cite{ilyushkin2017experimental}, all workloads are from the scientific domain and synthetically generated.
By using real-world traces from different domains, we investigate if differences are observed running these workloads, expanding prior knowledge in this field.

Our main result in this section is:
\begin{itemize}
	\item {\bf We find significant differences between autoscalers, \\when scheduling for different application domains.} 
	\item {\bf Autoscalers show different over- and underprovisioning behavior per workload.}
\end{itemize}

\subsection{Setup}

We use a real-world trace from the Chronos production environment at Shell and a real-world trace from the Askalon cluster.
The Chronos workload is composed of workflows that process \gls{iot} sensor data, sampled periodically.
The arrival pattern of this workload has an exponential pattern.
Every minute $2^i$ workflows arrive, where $i \in [0-9]$ corresponds to the amount of minutes into the workload.
As the sensors sample at a continuous rate, this workload can be repeated indefinitely.
The Askalon EE workload consists of workflows from the engineering domain.
The workload has a huge initial spike, roughly 16,000 tasks arrive in the first minute.
Overall Askalon EE spans 49 minutes.

Two setups are used in this experiment.
The first setup matches the state-of-the-art industrial setup described in \cite{ma2017ananke}.
This setup comprises 50 clusters with 70 resources (i.e. \glspl{vm}) each.
The other setup is tailored to each workload, where the number of clusters of the first setup is scaled to reach a system utilization of 70\%.

To measure the differences between autoscalers, we use the 
autoscaler-level metrics defined in Section~\ref{sssct:autoscaler-level_metrics}.

\subsection{Results}

The results of this experiment are visible in Table~\ref{tbl:workload_domain_experiment2}.
Overall, we observe significant differences in metrics per workload.
Some autoscalers severely overprovision on a workload, while it may underprovision on another.
This indicates that the aspects of the application domain, such as structure, arrival rate and complexity plays an important role on the performance of an autoscaler.

In particular, if we look at $A_U$, we notice that all autoscalers have similar values for all workloads.
The $A_O$, $M_u$, $\bar{V}$, $\bar{h}$, and $\bar{C}$ metrics for the Askalon EE workload are significantly different for Hist, ConPaaS and somewhat for Token.
This indicates that these autoscalers waste resources significantly more than others on this particular workload.
We ascribe these observations to the peak in Askalon EE.
Hist and ConPaaS keep a history of past arrival rates, effectively biasing future predictions due to this one time peak, while Token seems to overestimate the level of parallelism.

Overall, our results suggests the aspects that come with an application domain such as structure and complexity play an important role on the performance of an autoscaler.
This indicates the choice of autoscaler is non-trivial for a given application domain and should be carefully benchmarked.

\section{Bursty Workload Experiment}
\label{sct:bursty_workload_experiment}

Bursts (also referred to as peaks or flash-crowds) are common in cloud environments \cite{ali2014measuring}.
These sudden increases in demand of resources require a proactive approach to ensure task execution is not delayed.
The goal of this experiment is to provide insights to answer \emph{How well can autoscalers handle a significant burst in arriving tasks?}.

Our main result in this section is: 

\begin{itemize}
	\item \textbf{Workload agnostic autoscalers perform equally well compared to workload-specific autoscalers}.
	\item \textbf{The Plan autoscaler creates an order of magnitude more task delay than the other autoscalers for some workloads.}
\end{itemize}


\subsection{Setup}
\label{ssct:setup_bursty_workload_experiment}


\begin{figure}[!htb]
	\includegraphics[width=\columnwidth]{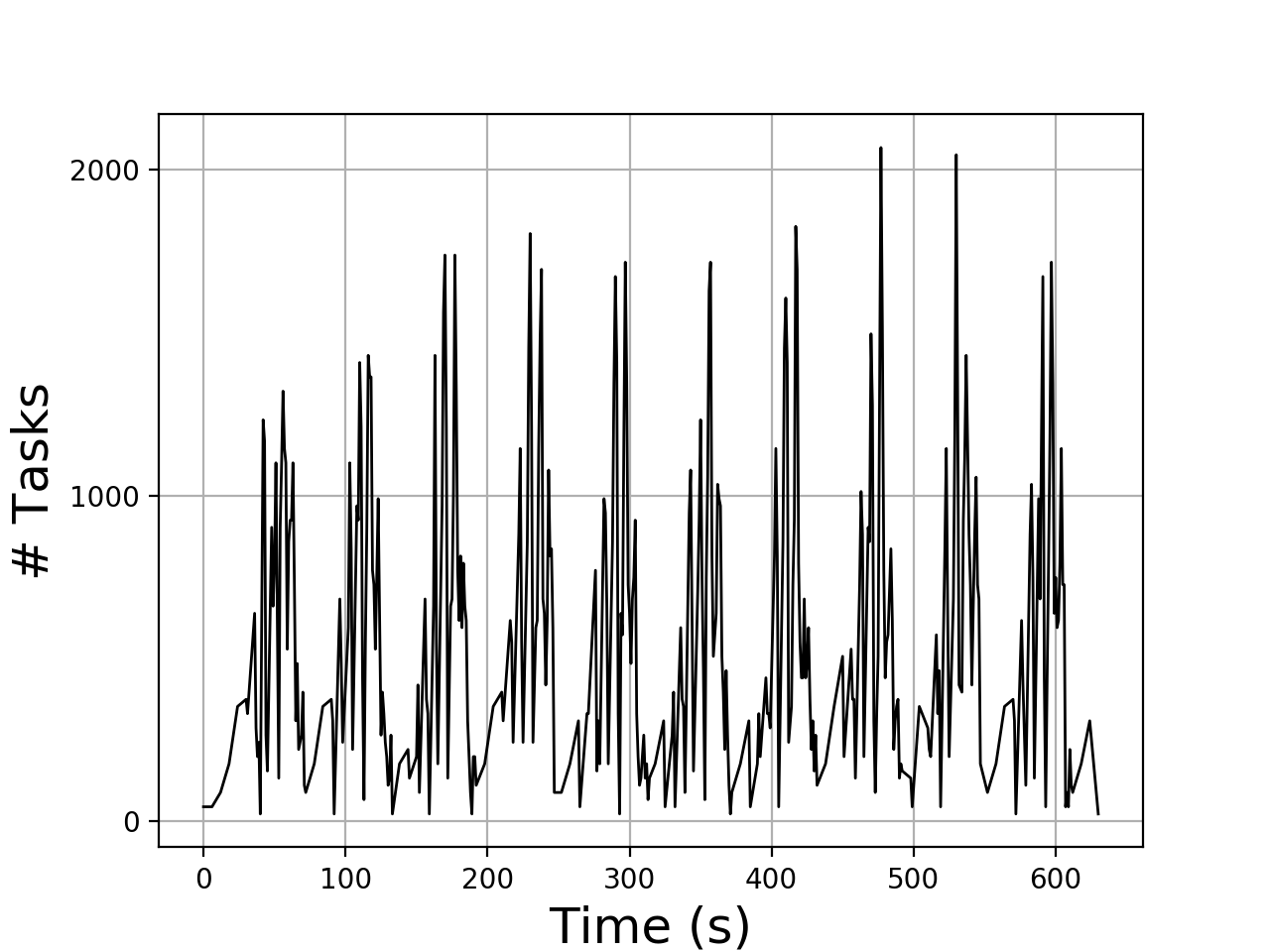}
	\caption{The arrival rate of Chronos (duplicated 22 times).}
	\label{fig:arrival_pattern_chronos_22x}
\end{figure}

To investigate the impact of bursts, we use two real-world workflows from two distinct domains.

The first workload is EE2 from the Askalon traces.
EE2 features an one-time burst at the start of the workload, visible in Figure~\ref{fig:arrival_pattern_askalon_ee2}.
In the first minute, around 24,000 tasks arrive which is not an uncommon amount in grids and clusters \cite{iosup2008grid}.

The second workload is a scaled version of the industrial Chronos workload used in Section~\ref{sct:different_domain_experiment}, visible in Figure~\ref{fig:arrival_pattern_chronos_22x}.
The workload is scaled so that the highest peak -- on a minute basis -- matches that of the Askalon EE2 trace.
This workload remains representative as the workload scales linearly with the amount of \gls{iot} sensors applied in Shell's Chronos infrastructure.

The infrastructure is designed to have an average of 70\% resource utilization, a representative number in supercomputing \cite{jones1999scheduling}.
To compute the number of sites needed to achieve this average utilization, we proceeded as follows.
First, we computed the total number of CPU seconds this workload is generating.
Next, we calculate the duration of the workload by computing the difference between the smallest submission time of all workflows and the latest completion time, using the critical path of the workflow as runtime.
Based on the amount of CPU seconds the workload generates and its duration, we compute the number of clusters required to obtain 70\% system utilization.
In these calculations we assume the sites run the entire duration of the workload and contain 70 \glspl{vm} (resources) each.
We calculate EE2 requires 13 sites and the scaled Chronos trace 62 sites to obtain an average 70\% utilization.

To measure the impact of bursts, we measure the cumulative delay for each workload per autoscaler and compute the normalized schedule length \cite{kwok1998benchmarking}.
The delay is computed by subtracting the critical path of a workload's makespan.
By using these metrics, we investigate if there are differences in how autoscalers handle burstiness.

\subsection{Results}

\begin{figure}[!htb]
	\hspace*{-0.5cm} 
	\centering 
	\includegraphics[width=1.1\columnwidth]{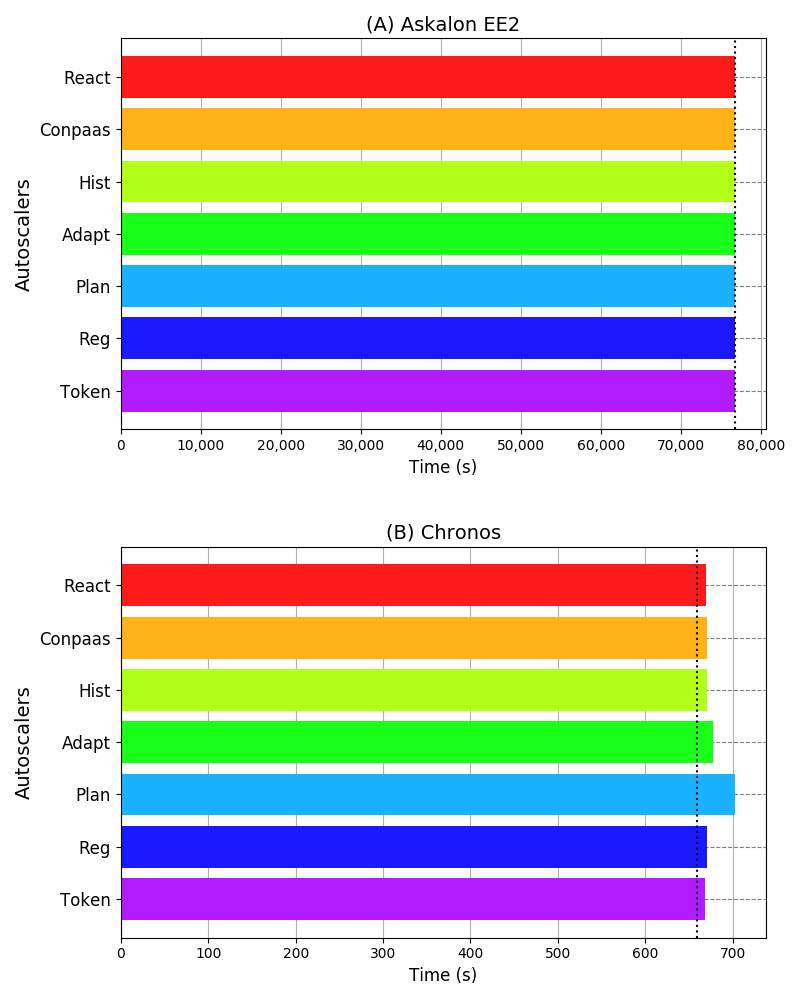}
	\caption[The makespan per autoscaler, per workload.]{The makespan per autoscaler, per workload. The vertical dotted line represents the critical path length of the workload. (Color-coding per autoscaler, matches Figure~\ref{fig:utilization_experiment_NSL_normalized_overprovisioning}.)}
	\label{fig:bursty_workload_experiment}
\end{figure}

\begin{figure}[!htb]
	\hspace*{-0.5cm} 
	\centering 
	\includegraphics[width=1.05\columnwidth]{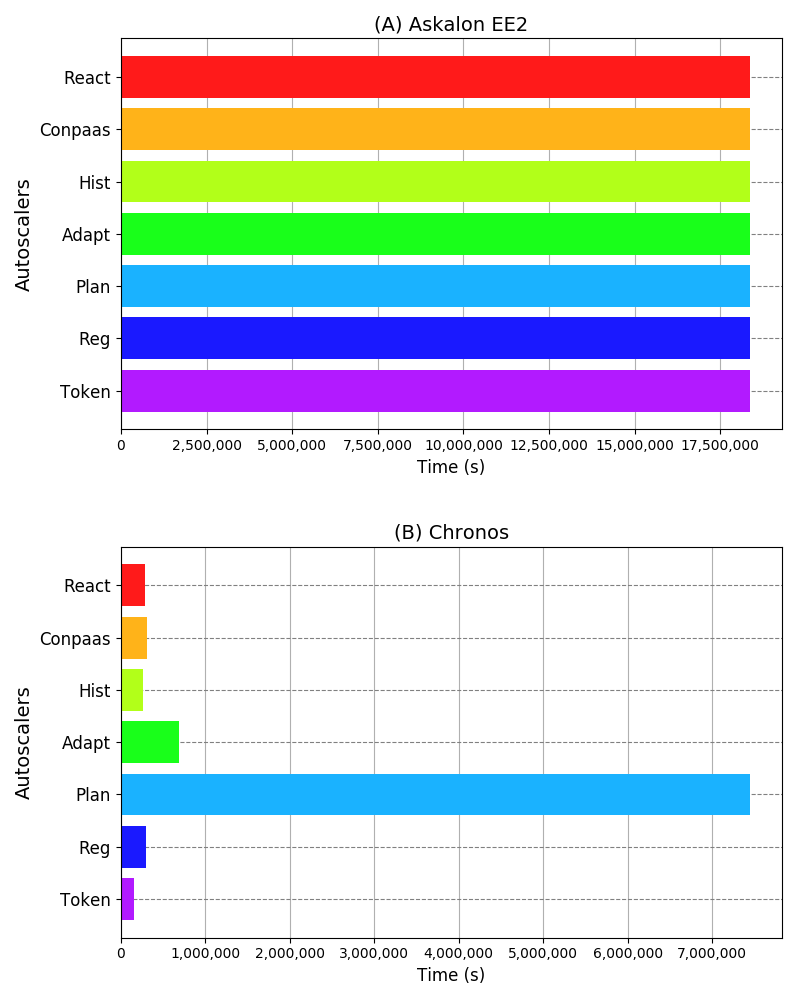}
	\caption{The cumulative delay per autoscaler, per workload.}
	\label{fig:cummulative_delay_bursty_workload_experiment}
\end{figure}

\begin{table}[!htb]
	\centering
	\caption{The NSL results, per autoscaler, per workload.}
	\label{tbl:bursty_workload_experiment_nsl}
	\resizebox{\columnwidth}{!}{%
	\begin{tabular}{|l||r|r|r|r|r|r|r|}
		\hline
		Workflow    & React & ConPaaS & Hist & Adapt & Plan & Reg  & Token \\ \hline \hline
        Askalon EE2 & 1.0   & 1.0     & 1.0  & 1.0   & 1.0  & 1.0  & 1.0  \\ \hline
        Chronos     & 1.02  & 1.02    & 1.02 & 1.03  & 1.07 & 1.02 & 1.01 \\ \hline
	\end{tabular}
}
\end{table}

To observe the impact of these bursts per autoscaler, we measure the M per workload and compute the overall NSL, per autoscaler.
Overall, there is a significant difference between the cumulative delay and M per workload.

The results are visible in Figure~\ref{fig:bursty_workload_experiment}.
This figure shows per autoscaler the M, per workload.
The CP of the workload is annotated by a black dotted line.
From this figure we conclude that the M is equal for the Askalon EE2 workload.
For the Chronos workload we observe Plan and Adapt have a slightly bigger M than the other autoscalers.
From the M and CP we compute the NSL for each combination, see Table~\ref{tbl:bursty_workload_experiment_nsl}.
From this table we observe significant differences in NSL between the two workloads, yet little difference between the autoscalers.

Figure~\ref{fig:cummulative_delay_bursty_workload_experiment} shows the cumulative delay per autoscaler per workload.
From this figure we observe that the delay is equal for all autoscalers when running the Askalon EE2 workload.
However, for the Chronos workload the Plan autoscaler creates an order of magnitude more delay than the other autoscalers.
This is in contrast with the NSL being roughly equal to the other autoscalers.
We therefore conclude that while the workload finishes in the same time compared to the other autoscalers, yet during execution tasks are delayed significantly more when using Plan.
This would likely lead to several \gls{qos} violations.

We thus conclude that differences can be observed between workloads in terms of NSL and autoscalers in terms of delay.
Additionally, the NSL does not tell the full story as the Plan autoscaler creates an order of magnitude more task delay executing Chronos while having a similar NSL compared to the other autoscalers.

\section{Different Allocation Policies Experiment}
\label{sct:different_allocation_policies_experiment}

\begin{figure}[t]
	\hspace*{-0.35cm} 
	\includegraphics[width=1.05\columnwidth]{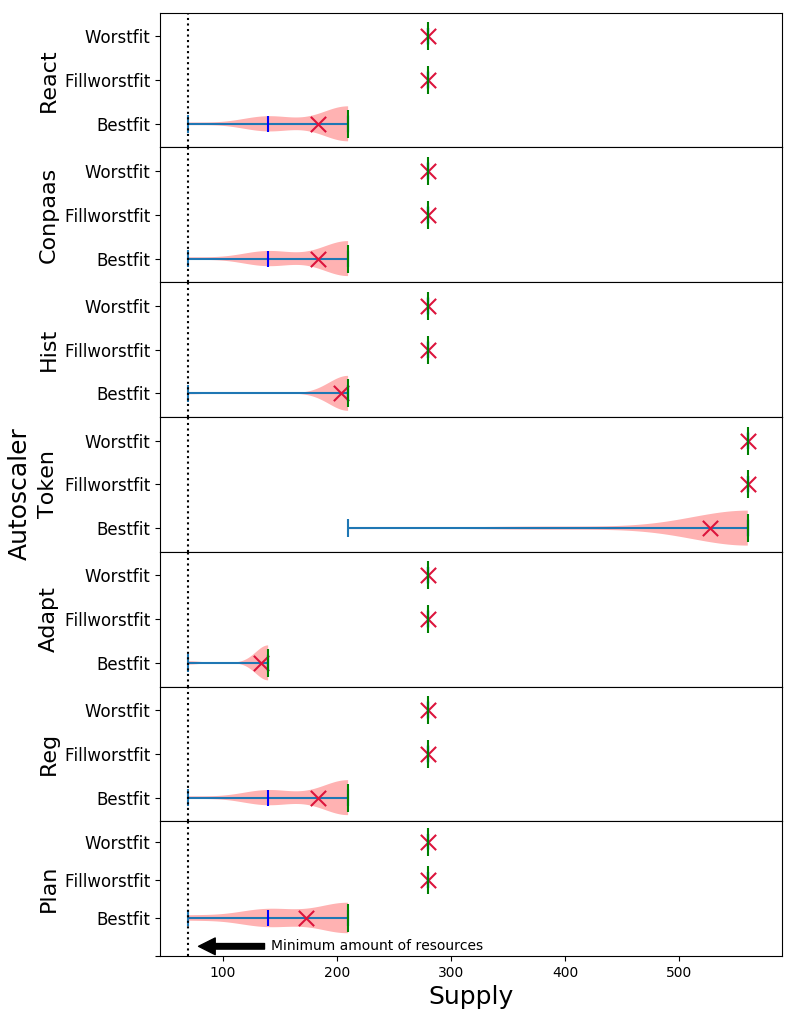}
	\caption[The supply in resources, per allocation policy, per provisioning policy.]{The supply in resources, per allocation policy, per provisioning policy. (For each allocation-autoscaler pair, the combined violin and box-and-whiskers plots summarize the empirical distribution of observed supply.)}
	\label{fig:allocation_experiment_result}
\end{figure}

Besides the availability of resources, other components can affect the performance of a system.
The goal of this experiment is to answer the research question \emph{Does the choice of allocation policy have an impact on the performance of the autoscalers?}, i.e. investigate and quantify the impact of the allocation policy on the performance of the autoscalers.

Our main result in this section is:
\begin{itemize}
	\item \textbf{The allocation policy has a direct influence on the behavior of all autoscalers.}
	\item \textbf{The allocation policy can have a significant impact on the average supply in a system.}
\end{itemize}

\subsection{Setup}

In this experiment we run all possible combinations of allocation policies and autoscalers using a state-of-the-art industrial setup and workload \cite{ma2017ananke}.

To measure the impact of allocation policies on the behavior of the autoscalers, we have implemented three different policies in our simulator: FillworstFit, WorstFit, and BestFit.

WorstFit selects the site with the most available resources.
After assigning a task it re-evaluates which site has the most available resources.
BestFit selects per task the site with the smallest number of available resources where the task still fits.
None of these policies use a greedy backfilling approach.

To measure the impact of proposed allocation policies, we measure the supply of resources throughout the experiment.

\subsection{Results}
\label{ssct:diff_allocation_experiment_results}

The results of this experiment are visible in Figure~\ref{fig:allocation_experiment_result}.
On the horizontal axis we show seven groups, one for each autoscaler.
On the vertical axis, we depict the supply computed by the autoscaler.
For each autoscaler and allocation policy combination, we show the distribution of the supply choices.
The minimum and maximum as well as the 25th and 75th percentiles are denoted by blue bars.
The median is denoted by a green bar.
The distribution of supply is visualized by a shaded area around the range.
The average supply is denoted by a cross symbol.

Overall we observe some differences between the autoscalers.
From the perspective of the autoscalers, besides Token, all autoscalers have the same minimum and maximum supply.

From the perspective of the allocation policies, we can observe the BestFit allocation policy has a significant lower average supply than the other two policies, for every autoscaler.
As FillworstFit and WorstFit assign jobs to the most idle clusters, clusters will rarely be considered idle and thus will not be deallocated.
One way to resolve this scenario is to migrate tasks by e.g. migrating the VM or interrupt and reschedule tasks.
This is part of our ongoing work.

From Figure~\ref{fig:allocation_experiment_result}, we observe adapt has the lowest maximum allocated resources, which impacts the response time of the workload.
All setups running FillworstFit and WorstFit require 650 seconds to process the workload.
Token running BestFit also requires 650 seconds as it provisions more machines.
Adapt running BestFit requires 694 seconds due to underprovisioning.
All other autoscalers running BestFit require 654 seconds.

From these observations we conclude that the allocation policy can have a direct influence on the behavior and performance of autoscalers.
As a result, comparing autoscaler can only be done fairly when using the same allocation policies.
\section{Different utilization experiment}
\label{sct:utilization_experiment}

\begin{table*}[] 
	\centering
	\caption{The scalability results, per autoscaler (AS), for different utilization levels ($U \in [0,1]$, $U=1$ for 100\% system load).}
	\label{tbl:autoscaler-level-experiments-utilization-exp}
	\begin{tabular}{|r||r|r|r|r|r|r|r|r|r|r|r|r|r|}
		\hline
		AS & Utilization & $A_U$ & $A_O$ & $\bar{A}_U$ & $\bar{A}_O$ & $T_U$ & $T_O$ & k & k' & $M_U$ & $\bar{V}$ & $\bar{h}$ & $\bar{C}$ \\ \hline
		\multirow{8}{*}{Reg} & 0.1 & 1.3 & 20.0 & 4.0 & 67.3 & 8.0 & 76.0 & 7.0 & 3.0 & 19.1 & 2,809.3 & 1,070.2 & 29.7\\
		& 0.2 & 24.3 & 14.9 & 10.9 & 53.3 & 19.0 & 65.0 & 6.0 & 2.0 & 16.3 & 1,733.1 & 1,310.7 & 36.4\\
		& 0.3 & 85.6 & 16.4 & 21.3 & 46.7 & 28.8 & 55.6 & 2.0 & 2.0 & 16.6 & 1,470.2 & 1,680.2 & 47.0\\
		& 0.5 & 317.5 & 35.2 & 39.6 & 44.8 & 46.7 & 49.7 & 1.0 & 1.0 & 35.9 & 1,614.8 & 3,075.8 & 86.0\\
		& 0.6 & 461.2 & 35.3 & 46.8 & 40.9 & 54.4 & 45.5 & 0.0 & 0.0 & 36.3 & 1,515.3 & 3,388.3 & 95.2\\
		& 0.7 & 597.4 & 32.1 & 51.9 & 36.0 & 59.3 & 39.8 & 0.0 & 0.0 & 33.2 & 1,332.8 & 3,427.1 & 102.8\\
		& 0.8 & 786.6 & 27.9 & 57.1 & 32.3 & 64.2 & 35.5 & 0.0 & 0.0 & 28.6 & 1,133.7 & 3,429.8 & 112.7\\
		& 0.9 & 958.5 & 27.9 & 60.7 & 29.5 & 67.1 & 32.8 & 0.0 & 0.0 & 28.8 & 1,032.2 & 3,539.1 & 126.0\\ \hline
		\multirow{8}{*}{Hist} & 0.1 & 1.3 & 80.4 & 1.0 & 89.3 & 5.0 & 94.0 & 22.0 & 0.0 & 80.0 & 8,515.5 & 3,244.0 & 90.1\\
		& 0.2 & 24.3 & 77.5 & 8.8 & 77.5 & 17.0 & 82.0 & 21.0 & 0.0 & 79.9 & 4,714.0 & 3,565.2 & 99.0\\
		& 0.3 & 86.3 & 68.5 & 19.5 & 68.5 & 27.0 & 72.0 & 17.0 & 0.0 & 69.8 & 3,119.5 & 3,565.2 & 99.0\\
		& 0.5 & 319.9 & 48.4 & 38.8 & 48.4 & 46.0 & 53.0 & 4.0 & 0.0 & 50.1 & 1,871.7 & 3,565.2 & 99.0\\
		& 0.6 & 461.2 & 41.0 & 46.8 & 41.0 & 54.4 & 45.5 & 0.0 & 0.0 & 42.9 & 1,607.3 & 3,594.1 & 101.0\\
		& 0.7 & 597.4 & 36.1 & 51.9 & 36.1 & 59.3 & 39.8 & 0.0 & 0.0 & 38.0 & 1,387.9 & 3,568.9 & 107.0\\
		& 0.8 & 786.6 & 32.4 & 57.1 & 32.4 & 64.2 & 35.5 & 0.0 & 0.0 & 33.8 & 1,187.0 & 3,590.9 & 118.0\\
		& 0.9 & 958.5 & 28.1 & 60.7 & 29.5 & 67.1 & 32.8 & 0.0 & 0.0 & 29.0 & 1,033.9 & 3,544.8 & 126.2\\ \hline
		\multirow{8}{*}{Conpaas} & 0.1 & 1.3 & 30.1 & 1.0 & 87.6 & 5.0 & 94.0 & 22.0 & 0.0 & 29.2 & 3,770.1 & 1,436.2 & 39.9\\
		& 0.2 & 24.3 & 47.0 & 8.8 & 76.7 & 17.0 & 82.0 & 21.0 & 0.0 & 48.4 & 3,265.2 & 2,469.5 & 68.6\\
		& 0.3 & 86.3 & 52.3 & 19.5 & 68.1 & 27.0 & 72.0 & 17.0 & 0.0 & 52.6 & 2,609.8 & 2,982.6 & 82.8\\
		& 0.5 & 319.9 & 45.5 & 38.8 & 48.4 & 46.0 & 53.0 & 4.0 & 0.0 & 46.2 & 1,817.1 & 3,461.2 & 96.1\\
		& 0.6 & 461.2 & 41.0 & 46.8 & 41.0 & 54.4 & 45.5 & 0.0 & 0.0 & 42.9 & 1,607.3 & 3,594.1 & 101.0\\
		& 0.7 & 597.4 & 36.1 & 51.9 & 36.1 & 59.3 & 39.8 & 0.0 & 0.0 & 38.0 & 1,387.9 & 3,568.9 & 107.0\\
		& 0.8 & 786.6 & 32.4 & 57.1 & 32.4 & 64.2 & 35.5 & 0.0 & 0.0 & 33.8 & 1,187.0 & 3,590.9 & 118.0\\
		& 0.9 & 958.5 & 28.8 & 60.7 & 29.5 & 67.1 & 32.8 & 0.0 & 0.0 & 30.2 & 1,041.0 & 3,569.1 & 127.1\\ \hline
		\multirow{8}{*}{React} & 0.1 & 1.3 & 20.1 & 3.0 & 73.2 & 7.0 & 82.0 & 12.0 & 2.0 & 19.2 & 2,817.7 & 1,073.4 & 29.8\\
		& 0.2 & 24.3 & 14.2 & 9.8 & 58.5 & 18.0 & 71.0 & 11.0 & 1.0 & 15.6 & 1,700.2 & 1,285.8 & 35.7\\
		& 0.3 & 86.3 & 15.9 & 20.5 & 53.9 & 28.0 & 63.0 & 9.0 & 1.0 & 16.2 & 1,462.1 & 1,671.0 & 46.4\\
		& 0.5 & 319.9 & 36.5 & 38.8 & 46.1 & 46.0 & 51.0 & 2.0 & 0.0 & 37.2 & 1,646.2 & 3,135.7 & 87.1\\
		& 0.6 & 461.2 & 36.3 & 46.8 & 40.9 & 54.4 & 45.5 & 0.0 & 0.0 & 37.3 & 1,531.2 & 3,423.9 & 96.2\\
		& 0.7 & 597.4 & 30.3 & 51.9 & 35.9 & 59.3 & 39.8 & 0.0 & 0.0 & 31.4 & 1,307.5 & 3,362.1 & 100.8\\
		& 0.8 & 786.6 & 29.7 & 57.1 & 32.3 & 64.2 & 35.5 & 0.0 & 0.0 & 30.4 & 1,155.6 & 3,496.0 & 114.9\\
		& 0.9 & 958.5 & 27.4 & 60.7 & 29.5 & 67.1 & 32.8 & 0.0 & 0.0 & 28.3 & 1,026.8 & 3,520.4 & 125.3\\ \hline
		\multirow{8}{*}{Token} & 0.1 & 1.3 & 20.2 & 1.0 & 85.8 & 5.0 & 94.0 & 22.0 & 0.0 & 19.3 & 2,833.8 & 1,079.6 & 30.0\\
		& 0.2 & 24.3 & 15.2 & 8.8 & 72.5 & 17.0 & 82.0 & 21.0 & 0.0 & 16.6 & 1,747.8 & 1,321.9 & 36.7\\
		& 0.3 & 86.3 & 16.9 & 19.5 & 63.0 & 27.0 & 72.0 & 17.0 & 0.0 & 17.2 & 1,493.6 & 1,707.0 & 47.4\\
		& 0.5 & 319.9 & 34.7 & 38.8 & 48.1 & 46.0 & 53.0 & 4.0 & 0.0 & 35.4 & 1,612.6 & 3,071.7 & 85.3\\
		& 0.6 & 461.2 & 35.7 & 46.8 & 40.9 & 54.4 & 45.5 & 0.0 & 0.0 & 36.7 & 1,522.2 & 3,403.8 & 95.7\\
		& 0.7 & 597.4 & 31.4 & 51.9 & 36.0 & 59.3 & 39.8 & 0.0 & 0.0 & 32.5 & 1,322.4 & 3,400.4 & 102.0\\
		& 0.8 & 786.6 & 29.0 & 57.1 & 32.3 & 64.2 & 35.5 & 0.0 & 0.0 & 29.7 & 1,146.7 & 3,469.1 & 114.0\\
		& 0.9 & 958.5 & 27.0 & 60.7 & 29.5 & 67.1 & 32.8 & 0.0 & 0.0 & 27.9 & 1,023.0 & 3,507.3 & 124.9\\ \hline
		\multirow{8}{*}{Adapt} & 0.1 & 1.3 & 19.9 & 3.0 & 73.2 & 7.0 & 82.0 & 12.0 & 2.0 & 19.0 & 2,804.4 & 1,068.4 & 29.7\\
		& 0.2 & 24.3 & 14.5 & 9.8 & 58.5 & 18.0 & 71.0 & 11.0 & 1.0 & 16.0 & 1,718.4 & 1,299.6 & 36.1\\
		& 0.3 & 86.3 & 15.7 & 20.5 & 53.9 & 28.0 & 63.0 & 9.0 & 1.0 & 16.0 & 1,455.8 & 1,663.8 & 46.2\\
		& 0.5 & 319.9 & 37.8 & 38.8 & 46.2 & 46.0 & 51.0 & 2.0 & 0.0 & 38.5 & 1,670.8 & 3,182.4 & 88.4\\
		& 0.6 & 461.2 & 33.0 & 46.8 & 40.8 & 54.4 & 45.5 & 0.0 & 0.0 & 34.0 & 1,478.6 & 3,306.3 & 92.9\\
		& 0.7 & 597.4 & 30.9 & 51.9 & 35.9 & 59.3 & 39.8 & 0.0 & 0.0 & 31.9 & 1,315.3 & 3,382.1 & 101.4\\
		& 0.8 & 786.6 & 29.1 & 57.1 & 32.3 & 64.2 & 35.5 & 0.0 & 0.0 & 29.8 & 1,148.5 & 3,474.5 & 114.2\\
		& 0.9 & 958.5 & 27.1 & 60.7 & 29.5 & 67.1 & 32.8 & 0.0 & 0.0 & 28.0 & 1,023.5 & 3,509.2 & 124.9\\ \hline
		\multirow{8}{*}{Plan} & 0.1 & 1.3 & 19.6 & 5.0 & 63.7 & 8.9 & 72.5 & 4.0 & 4.0 & 18.7 & 2,765.9 & 1,053.7 & 29.5\\
		& 0.2 & 24.1 & 14.2 & 11.8 & 48.8 & 19.9 & 61.6 & 3.0 & 3.0 & 15.5 & 1,692.0 & 1,279.6 & 35.8\\
		& 0.3 & 85.6 & 15.2 & 21.3 & 46.3 & 28.8 & 55.6 & 2.0 & 2.0 & 15.5 & 1,434.1 & 1,638.9 & 45.8\\
		& 0.5 & 317.5 & 34.9 & 39.6 & 44.8 & 46.7 & 49.7 & 1.0 & 1.0 & 35.6 & 1,609.9 & 3,066.5 & 85.8\\
		& 0.6 & 461.2 & 33.2 & 46.8 & 40.8 & 54.4 & 45.5 & 0.0 & 0.0 & 34.2 & 1,481.4 & 3,312.5 & 93.1\\
		& 0.7 & 597.4 & 32.7 & 51.9 & 36.0 & 59.3 & 39.8 & 0.0 & 0.0 & 33.8 & 1,341.2 & 3,448.8 & 103.4\\
		& 0.8 & 786.6 & 28.9 & 57.1 & 32.3 & 64.2 & 35.5 & 0.0 & 0.0 & 29.5 & 1,145.6 & 3,465.6 & 113.9\\
		& 0.9 & 958.5 & 26.6 & 60.7 & 29.4 & 67.1 & 32.8 & 0.0 & 0.0 & 27.5 & 1,018.0 & 3,490.5 & 124.3\\ \hline
		
	\end{tabular}
\end{table*}

\begin{figure}[h]
	\centering 
	\includegraphics[width=1.1\columnwidth]{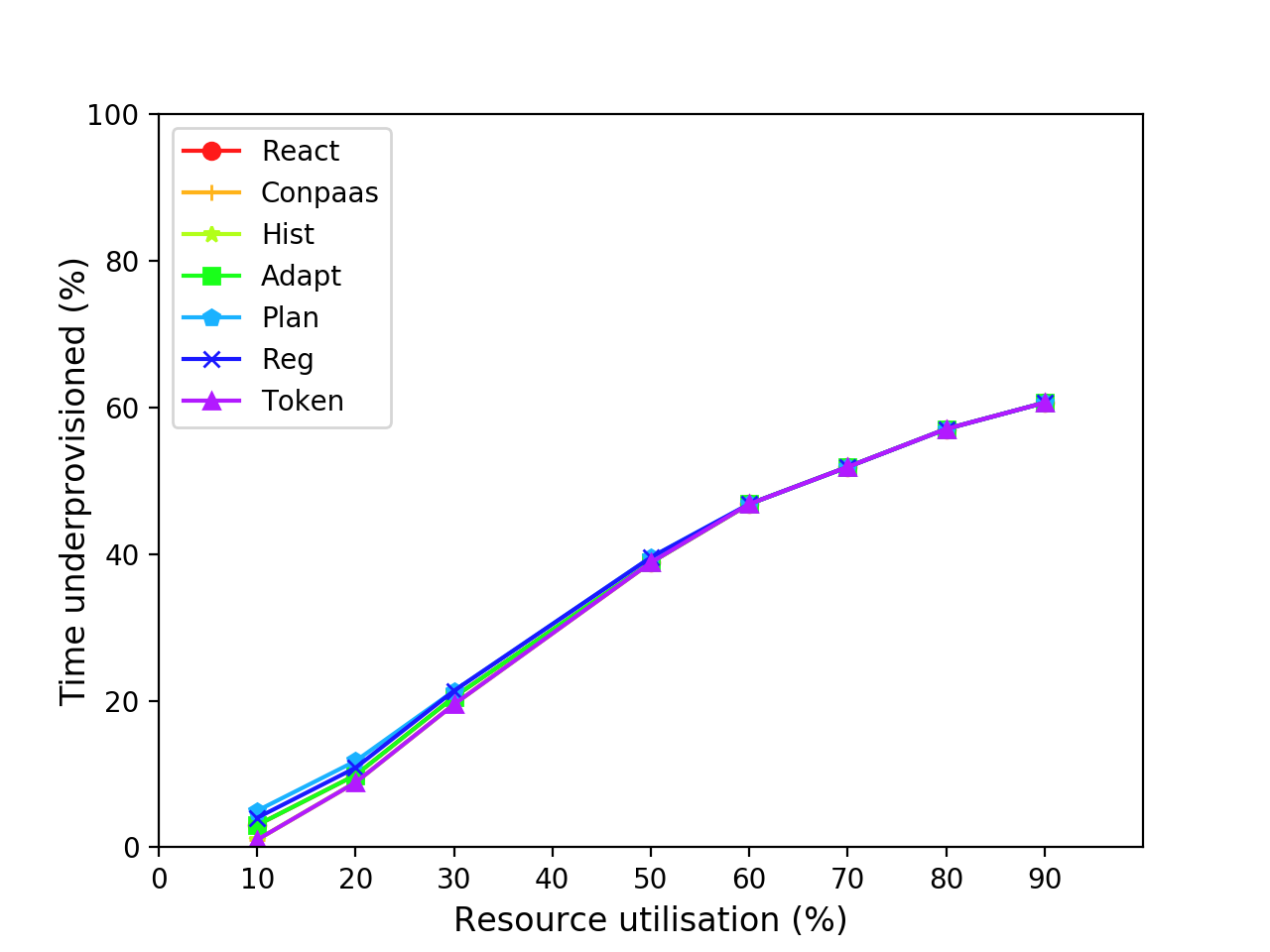}
	\caption{The normalized underprovisioning per utilization per autoscaler.}
	\label{fig:utilization_experiment_normalized_underprovisioning}
\end{figure}

\begin{figure}[h]
	\hspace*{-0.3cm}
	\centering 
	\includegraphics[width=1.05\columnwidth]{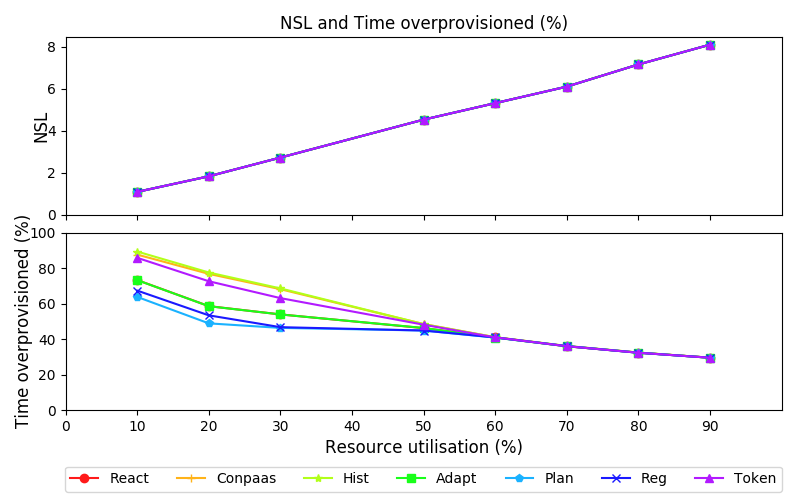}
	\caption{The NSL and normalized overprovisioning per utilization per autoscaler.}
	\label{fig:utilization_experiment_NSL_normalized_overprovisioning}
\end{figure}

In the current state-of-the-art \cite{ilyushkin2017experimental}, the system utilization is 39.5\% for workload 1, assuming all \glspl{vm} were allocated throughout the experiment.
Prior work demonstrates a resource utilization of 70\% is possible \cite{jones1999scheduling} in supercomputing.
In this experiment, we shed light on our research question \emph{How do autoscalers behave when faced with different resource environments?}.
To investigate how autoscalers operate when faced with different amount of resources, we scale the infrastructure based on a target resource utilization based on the amount of CPU seconds a workload contains.
We compare the normalized under- and overprovisioning for each autoscaler.

Our main result in this section is: 

\begin{itemize}
	\item \textbf{All autoscalers underprovision roughly the same fraction of time.}
	\item \textbf{The fraction of time of overprovisioning differs per autoscaler significantly at lower utilizations, yet converges for high utilizations.}
	\item \textbf{Autoscalers that significantly overprovision more resources do not yield a better NSL.}
\end{itemize}

\subsection{Setup}
In this experiment we run the Askalon EE workload, a workload from the engineering domain.
This workload contains a load of 2,823,758 CPU seconds and spans in total 2,998 seconds, based on the submission times and critical paths of the workflows.
We vary the number of clusters to obtain 10\%, 20\%, \dots, 90\% resource utilization, based on the amount of CPU seconds and span of the workload.
Each cluster contains 70 \glspl{vm}.
Note that since autoscalers predict the amount of \glspl{vm} required, the system may allocate more \glspl{vm} since the resource manager computes the required amount of clusters (ceiled) to meet the prediction (up to the maximum amount of clusters specified).

We measure the normalized under- and overprovisioning time in the system as well as the normalized schedule length (NSL) \cite{kwok1998benchmarking} of processing the workload.
This provides us with insight into the trade-offs autoscalers make when allocating resources.

\subsection{Results}

The results are visible in Figures~\ref{fig:utilization_experiment_normalized_underprovisioning} and \ref{fig:utilization_experiment_NSL_normalized_overprovisioning}.
In these figures we plot for each resource utilization percentage the $\bar{A}_U$, and $\bar{A}_O$ and NSL.

From Figure~\ref{fig:utilization_experiment_normalized_underprovisioning} we observe no significant differences in $\bar{A}_U$ between the autoscalers.
Figure~\ref{fig:utilization_experiment_NSL_normalized_overprovisioning} does show significant differences between the autoscalers for $\bar{A}_O$.

We believe the reason for $\bar{A}_U$ being almost identical for all autoscalers lies in the arrival pattern of the EE workload.
Due to the burst at the start, the autoscalers predict a high amount of resources needed.
This amount can be provisioned when running in an environment having a low resource utilization.
As the amount of resources available is reduced to achieve a higher resource utilization, the $\bar{A}_U$ increases as the amount of resources cannot be allocated according to the prediction.

$\bar{A}_O$ shows significant differences between the autoscalers.
We expect autoscalers to somewhat overprovision, as resources are incremented per 70, and the system always overprovisions if it cannot meet the target exactly.
However, we observe Hist, ConPaaS, and Token overprovision significantly more than the other autoscalers, while yielding no benefit in NSL.
We ascribe this behavior for Hist and ConPaaS to keeping a history of incoming jobs.
For EE, a lot jobs arrive at the start, causing the histogram to be biased towards overprovisioning resources.
For Token, the level of parallelism influences the amount of resources provisioned.
Token estimates a high level of parallelism for the EE workload, causing overprovisioning.

From these observations we conclude that all autoscalers roughly underprovisions the same fraction of time.
At lower system utilization, some autoscalers overprovision more than others, yet converges for high system utilization.
The autoscalers that overprovision more resources do not yield a better NSL.

The results for all autoscaler-level metrics are in Table~\ref{tbl:autoscaler-level-experiments-utilization-exp}.
From this table we observe a higher $k$ for Hist, ConPaaS, and Token, which corresponds to our overprovisioning findings.
Correspondingly, $T_O$, $\bar{h}$, and $C$ are also higher for these three autoscalers.
In particular Hist seems to be wasting most resources as $M_u$ and $C$ are significantly higher than all other autoscalers.


\section{Related Work}\label{sec:related}

This study complements, and by design of research questions significantly extends, the large body of related work in the field surveyed by Lorido-Botran et al.~\cite{lorido2014review} and by Vaquero et al.~\cite{vaquero2011dynamically}. Beyond the new research questions we propose, our study is the first to use an experimental approach with workloads from multiple domains, and diverse datacenter environments.  


Closest to our work, Ilyushkin et al.~\cite{ilyushkin2017experimental} conduct the first comprehensive comparison of autoscalers for workflows, using real-world experimentation. The experimental design is limited: their work focuses on the scientific domain, and, due to the use of real-world resources, limits the amount of machines to 50 and the resources used per task to a single core.


Also close to this work, Papadopoulos et al.~\cite{papadopoulos2016peas} introduce a performance evaluation framework for autoscaling strategies in cloud applications (PEAS).
PEAS measures the performance of autoscaling strategies using scenario theory. Similarly to this work, they use a simulation-based experimental environment, analyzing the elasticity of 6 autoscaling policies for workload traces collected from the web-hosting domain. In contrast, this work focuses on workloads of {\it workflows}, leading to significantly new scheduling constraints, on different application domains, and on different types of metrics (in particular, also workflow-level).

Our work also complements the specialized performance comparisons presented by authors of new autoscalers, such as the work of Han et al.~\cite{han2012lightweight} (domain-agnostic), Hasan et al.~\cite{hasan2012integrated} (domain-agnostic), Heinze et al.~\cite{heinze2014auto} (stream-processing domain), Mao et al.~\cite{mao2011auto} (scientific workflows), Jiang et al.~\cite{jiang2013optimal} (web-hosting domain), Dougherty et al.~\cite{dougherty2012model} (focus on energy-related metrics). In contrast to these studies, ours focuses on deeper analysis focusing on new research questions, on different and more diverse application domains, etc.

\section{Threats to Validity}

The limitations of this work mainly embodies itself in the use of the simulator.
To combat this threat, we have validated our simulator using small example workflows, generated workflows using Little's Law and reran the experiments defined in \cite{ilyushkin2017experimental}.
The use of real-world traces in all experiments is to further improve the resemblance to the real-world.

The simulated infrastructure could be improved upon.
In this work, we assume no (or negligible) IO time.
However, as workflows can be both compute- and IO-intensive \cite{may2006zib}, modeling IO  increases the representativeness of the results as it's becoming increasingly important \cite{eom2000speed}. 

Another aspect is all experiments using a homogeneous setup.
Private/hybrid clouds such as the DAS~\cite{bal2016medium} are often homogeneous~\cite{ghanbari2012feedback} or provide a homogeneous view of virtualized resources \cite{sotomayor2009virtual}.
While related work such as \cite{ma2017ananke} and \cite{ilyushkin2017experimental} use homogeneous setups, investigating the effect of heterogeneity could further improve the results in this paper and is part of our ongoing work.

One other improvement regarding resource usage is to measure network usage.
Network usage is not a critical factor in job completion time \cite{ousterhout2015making}, yet can be used to improve performance \cite{uta2016towards}.
Ideally, a real-world setup could be capturing such metrics to further classify and compare the autoscalers.

The domains used in this work do not represent the full spectrum found in cloud environments.
Workloads from e.g. health, financial and business domains could give further insight on the applicability of the autoscalers.
In this work we covered the scientific, engineering and industrial domains, yet demographic statistics show a significant amount of cloud use comes from other domains such as the financial sector \cite{garrison2012success}.
Experimenting with more distinct domains and multiple workloads per domain will provide deeper insight into the workings of the autoscalers and differences per domain.
\section{Conclusion and Ongoing Work}
\label{sct:conclusion}\label{sec:conclusion}\label{sec:futurework}

Autoscaling, that is, automating the process of acquiring and releasing resources at runtime, is an important, non-trivial task at the core of datacenter operation.
To help with understanding how autoscalers work, prior work has performed systematic analysis and comparisons, yet the results still lack in the diversity of application domain, choice of metrics, and environments.
Addressing this lack of knowledge, in this work we perform a comprehensive comparison of state-of-the-art autoscalers. 

We ask new research questions about the operational laws of autoscalers managing workloads of workflows. To answer these questions, we conduct trace-based simulations, measuring the impact of autoscaling across a variety of datacenter environments, workloads, and metrics. We use workloads from three different domains, scientific, industrial processes, and engineering. We analyze the performance effects of workload burstiness, of the interplay between allocation and autoscaling, and of the level of utilization in the datacenter. 


Our study gives strong, quantitative evidence about autoscaling performance, including findings such as:
\begin{itemize}
	\item The application domain has a direct impact on the performance of an autoscaler.
	\item For bursty workloads, workload agnostic autoscalers perform equally well compared to workfload-specific autoscalers in terms of NSL.
	\item The allocation policy has a direct impact on the performance of an autoscaler.
	\item Some autoscalers overprovision more than others, while yielding no benefit to workflow-level metrics such as NSL.
\end{itemize}

In our ongoing work, we will study the impact of heterogeneity in datacenters on autoscalers, measure the impact of different application domains, using traditional and emerging cost metrics (e.g., the finer-grained cost-models released for selected resources by Microsoft and Google in mid-2017) to compare autoscalers, etc. 
As described in Section~\ref{ssct:diff_allocation_experiment_results}, we will investigate the effect of migrating jobs in order to deallocate clusters, using e.g. VM migrations or interrupting tasks, to gain further insight into the effect of using different allocation policies.
%

\bibliographystyle{IEEEtran.bst}
\bibliography{bibliography.bib} 

\end{document}